\documentclass[a4paper,conference]{IEEEtran}

\usepackage[left=1.57cm,right=1.57cm,top=0.95cm,bottom=2.54cm]{geometry}

\usepackage[american]{babel}
\usepackage{hyphenat}

\usepackage{algorithmic}
\usepackage{algorithm}
\usepackage{array}

\usepackage{subfigure}

\usepackage{textcomp}
\usepackage{url}
\usepackage{verbatim}
\usepackage{graphicx}
\usepackage{xcolor}

\usepackage{cite}
\usepackage[utf8]{inputenc}
\usepackage[T1]{fontenc}
\usepackage{textcomp}
\usepackage{microtype}
\usepackage{calc}
\usepackage[normalem]{ulem}

\usepackage{amsmath}
\usepackage{amssymb}
\usepackage{amsfonts}
\usepackage{amsthm}

\usepackage{mathtools}

\usepackage[capitalize,noabbrev,nameinlink]{cleveref}

\RequirePackage{xstring}
\RequirePackage{xparse}
\RequirePackage[]{acro}
\makeatletter
\@ifpackagelater{acro}{2019/02/17}{
	\NewDocumentCommand\acrodef{mO{#1}mG{}}{\DeclareAcronym{#1}{short={#2}, long={#3}, foreign-plural={}, #4}}
}{
	\NewDocumentCommand\acrodef{mO{#1}mG{}}{\DeclareAcronym{#1}{short={#2}, long={#3}, #4}}
}
\makeatother
\acrodef{AIFS}{Arbitration Inter-Frame Spacing}
\acrodef{AoI}{Age of Information}
\acrodef{AWGN}{Additive White Gaussian Noise}
\acrodef{BS}{Base Station}
\acrodef{BSS}{Basic Service Set}
\acrodef{CAM}{Cooperative Awareness packet}
\acrodef{CARD}{Controlled Asymmetry Reduces Delay}
\acrodef{CBF}{Contention-Based Forwarding}
\acrodef{CBR}{Channel Busy Ratio}
\acrodef{CCDF}{Complementary Cumulative Distribution Function}
\acrodef{CCH}{Control Channel}
\acrodef{CDMA}{Code Division Multiple Access}
\acrodef{CDF}{Cumulative Distribution Function}
\acrodef{C-ITS}{Cooperative Intelligent Transportation System}{short-plural-form={C-ITS}}
\acrodef{COV}{Coefficient Of Variation}
\acrodef{CPM}{Collective Perception packet}
\acrodef{CPU}{Central Processing Unit}
\acrodef{DSRC}{Dedicated Short-Range Communication}
\acrodef{DTMC}{Discrete Time Markov Chain}
\acrodef{FCFS}{First Come First Served}
\acrodef{GFMA}{Grant Free Multiple Access}
\acrodef{GNCU}{Google Normalized Computing Unit}
\acrodef{GNMU}{Google Normalized Memory Unit}
\acrodef{i.i.d.}{independent and identically distributed}
\acrodef{IAT}{Inter arrival times}
\acrodef{JIQ}{Join Idle Queue}
\acrodef{JSQ}{Join Shortest Queue}
\acrodef{LCFS}{Last Come First Served}
\acrodef{LCFSwO}{Last Come First Served with Overwrite}
\acrodef{LDM}{Local Dynamic Map}
\acrodef{LWL}{Least Work Left}
\acrodef{mMTC}{Massive machine-type Communication}
\acrodef{M2M}{Machine-to-Machine}
\acrodef{MAC}{Medium Access Control}
\acrodef{MCM}{Maneuver Coordination packet}
\acrodef{MCS}{Modulation and Coding Scheme}
\acrodef{MPR}{Multi-Packet Reception}
\acrodef{MRT}{Mean Response Time}
\acrodef{MUSA}{Multi User Shared Access}
\acrodef{NOMA}{Non-Orthogonal Multiple Access}
\acrodef{NoB}{No Buffering}
\acrodef{OBU}{On-Board Unit}
\acrodef{OMA}{Orthogonal Multiple Access}
\acrodef{PDF}{Probability Density Function}
\acrodef{PDMA}{Pattern Division Multiple Access}
\acrodef{PER}{Packet Error Rate}
\acrodef{PSJF}{Preemptive-Shortest-Job-First}
\acrodef{RAM}{Random Access Memory}
\acrodef{RR}{Round Robin}
\acrodef{RSU}{Road Side Unit}
\acrodef{SCH}{Service Channel}
\acrodef{SCOV}{Squared Coefficient Of Variation}
\acrodef{SA}{Slotted ALOHA}
\acrodef{SD}{Slowdown}
\acrodef{SIC}{Successive Interference Cancellation}
\acrodef{SITA}{Size Interval Task Assignment}
\acrodef{SNR}{Signal to Noise Ratio}
\acrodef{SNIR}{Signal to Noise plus Interference Ratio}
\acrodef{SRPT}{Shortest Remaining Processing Time}
\acrodef{TRG}{Two-Ray Ground}
\acrodef{V2I}{Vehicle-to-Infrastructure}
\acrodef{V2V}{Vehicle-to-Vehicle}
\acrodef{V2X}{Vehicle-to-Everything}
\acrodef{VANET}{Vehicular Ad Hoc Network}
\acrodef{VLC}{Visible Light Communication}
\acrodef{UAV}{Unmanned Aerial Vehicle}
\acrodef{JBT}{Join Below Threshold}

\def\todoCtd#1{%
	TODO: #1%
	\ifx&#1&.\fi%
	\endgroup
	\cbend
	\relax
}

\begin{document}

\title{``Two-Stagification'': Job Dispatching in Large-Scale Clusters via a Two-Stage Architecture}




\author{%
\IEEEauthorblockN{%
	Mert Yildiz\IEEEauthorrefmark{1}%
	,
	Alexey Rolich\IEEEauthorrefmark{1}%
	,
	Andrea Baiocchi\IEEEauthorrefmark{1}%
}%

\IEEEauthorblockA{
	\IEEEauthorrefmark{1}\small{Dept.\ of Information Engineering, Electronics, and Telecommunications (DIET), University of Rome Sapienza, Italy}
}%


%
\texttt{%
	\{mert.yildiz,alexey.rolich,andrea.baiocchi\}%
	@uniroma1.it%
}%
}

\maketitle

\begin{abstract}\nohyphens{%
A continuing effort is devoted to devising effective dispatching policies for clusters of First Come First Served servers.
Although the optimal solution for dispatchers aware of both job size and server state remains elusive, lower bounds and strong heuristics are known. 
In this paper, we introduce a two-stage cluster architecture that applies classical Round Robin, Join Idle Queue, and Least Work Left dispatching schemes, coupled with an optimized service-time threshold to separate large jobs from shorter ones. 
Using both synthetic (Weibull) workloads and real Google data center traces, we demonstrate that our two-stage approach greatly improves upon the corresponding single-stage policies and closely approaches the performance of advanced size- and state-aware methods.
Our results highlight that careful architectural design—rather than increased complexity at the dispatcher—can yield significantly better mean response times in large-scale computing environments.
}\end{abstract}


\acresetall
\IEEEpeerreviewmaketitle

%

\section{Introduction}
\label{sec:introduction}

Finding effective and easily implementable algorithms for distributing the workload in server clusters is among the top issues in the framework of cloud and edge computing.
Given a set of servers to which a sequence of jobs is offered, possibly broken down into several tasks, dispatching policies assign each job or task to one server.
Dispatching is therefore a workload distribution function, unlike scheduling, which is implemented \emph{inside} each server and defines when and how much of the processing facility of the server is assigned to each job or task dispatched to that server.

Dispatching decisions are typically assumed to be immediate, i.e., upon job arrival.
Dispatching policies can be classified according to the amount of information available at the dispatching process.
The simplest policies are stateless and non-anticipative, i.e., no information is available on the current state of servers, nor are job/task sizes known at time of dispatching.
In this framework, we distinguish between memoryless strategies (e.g., random selection of a server), or strategies that keep a memory of past choices (e.g., \ac{RR}).
Stateful dispatching policies base their decisions on knowledge of the current system state, e.g., by means of server polling.
\ac{JIQ} and \ac{JSQ} are examples of this class of policies.
Finally, anticipative policies (also known as size-aware policies) exploit knowledge of job/task size upon arrival, before the job/task is run.
\ac{LWL} and \ac{SITA} are classic examples of this kind of policies.
Other classification criteria are possible, e.g., based on time variability of policy parameters.
Static policies maintain values of their parameters unchanged over time, while dynamic policies adapt their parameters and rules, e.g., based on running time estimates of relevant quantities.

While many dispatching and scheduling policies have been defined and analyzed under different assumptions (e.g., see \cite{HarcholBalter2013,Zhou2017,KUMAR2019,Tang2024}), still many open issues are on the table \cite{HarcholBalter2021}.
We focus on a server cluster made up of $n$ identical servers that serve jobs according to \ac{FCFS} scheduling.
A full understanding of even such a simple multi-server cluster system is yet to be achieved.
The optimal dispatching policy is known in some cases, e.g., round robin is optimal under the assumption that nothing is known to the dispatching process, except for the history of previous dispatching decisions; \ac{SITA} is optimal when job sizes are known upon arrival, but no server state information is available.
However, to date, the exact optimal policy in the general case, where job size is known upon arrival as well as the current status of servers, is not characterized, even in the relatively simple M/G-type model setting (i.e., Poisson arrival of jobs, renewal job sizes).

On the other hand, several powerful heuristics have been proposed.
Dynamic sequential policies are defined in \cite{Hyytia2022}.
The key idea is to start from a baseline policy, to define a sequential ordering of servers.
Servers are then polled according to the considered sequential order, checking a condition, that can depend on server status and on job size. 
The job is assigned to the first server that matches the condition.
The optimal policy is derived numerically in \cite{Hyytia2024} from Bellman-type optimality equations in a cluster of servers with Poisson job arrivals, then generalized to renewal inter-arrival times.
This approach however calls for significant complexity with growing server cluster size.
A general lower bound of \ac{MRT} for the class of stateful, anticipative dispatching policies is established in \cite{Xie2024}.
A heavy-traffic delay optimal heuristic policy is defined as well, named \ac{CARD}\cite{Xie2024}.
A version of this algorithm, referred to as multi-band flexible \ac{CARD}, is defined for arbitrary server cluster size.
This policy is shown to yield very good performance at every load level, not just asymptotically in heavy traffic.

A common trait of proposed heuristic policies is their complexity.
They require knowledge of job size upon arrival, and knowledge of current server status, which practically requires message exchange between the dispatching process and servers.
An effort towards simpler yet effective dispatching policies was the motivation of \cite{Zhou2017}, where \ac{JBT} is defined and proved to be heavy-traffic delay optimal.
This policy is a generalization of the sub-optimal \ac{JIQ} policy.
The dispatcher keeps track of servers whose backlog is less than $r$ jobs (the threshold) and jobs are assigned to one of those servers, if available, randomly to any server otherwise.
The dynamic version of this policy adapts the threshold according to an estimate of the current backlog of servers.
The estimate is obtained by means of polling of a few randomly selected servers.

The core problem addressed in this work is designing effective, implementable dispatching policies for \ac{FCFS} server clusters. While advanced state- and size-aware strategies exist, the optimal policy remains unknown and existing heuristics tend to be complex. The motivation is to assess whether simple, low-overhead policies—requiring minimal system knowledge—can match the performance of more sophisticated methods in large-scale cloud and edge environments.

In this paper, we explore how to achieve comparably good performance in terms of \ac{MRT} by means of very simple dispatching policies.
Our key contributions are:
\begin{itemize} 
\item Demonstrating that simple, size- and state-agnostic policies (e.g., \ac{RR}) can match the performance of more complex strategies. 
\item Showing that a two-stage server architecture significantly improves performance over traditional single-stage dispatching, benefiting not only \ac{RR} but also \ac{JIQ} and \ac{LWL}.
\end{itemize}

In the two-stage system, jobs are simply offered to the first stage of servers upon their arrival.
If a job is not completed within a given time threshold, it is stopped, moved to the next stage, and restarted.
We show that it is possible to improve over size-aware best heuristics, provided the threshold and the sizes of the two stages are properly optimized.
While our current analysis does not delve into the overhead potentially introduced by the job transfer between stages, we acknowledge its relevance. This aspect will be addressed in future work, where we plan to quantify its impact on overall system performance.

To that end, we first consider an artificially generated workload, based on Poisson arrivals and renewal job sizes, according to the classic M/G-type model setting.
To confirm obtained results in a more realistic environment, we exploit high-quality workload measurements of production data centers, specifically those published recently by Google \cite{wilkes2019clusterdata, Tirmazi2020, Sfakianakis2021Tracebased, yildiz2024, yildiz2025meritsimplepoliciesbuying, yildiz2025dispatchingodysseyexploringperformance}. 
In that dataset, job arrivals do not follow a Poisson process, and the job/task sizes do not adhere to a negative exponential distribution \cite{yildiz2024}. 
On the contrary, the distribution of service times is highly variable and exhibits an extremely heavy-tailed behavior, whereby a small fraction of the very largest jobs comprises most of the load. 
In prior empirical studies of computational workloads  \cite{crovella1998,harcholbalter1996,harcholbalter1996Downey,harcholbalter1999,harcholbalter2003}, the authors found that the top 1\% of jobs comprise 50\% of the load.
The heavy-tailed property exhibited in Google data centers is even more extreme, namely, the largest (most compute-intensive) 1\% of jobs comprise about 99\% of the load \cite{Tirmazi2020}.
This makes performance evaluation based on Google's measured workload particularly challenging for dispatching policies.

The rest of the paper is organized as follows.
\cref{sec:modeldescription} provides a brief explanation of the considered system model and workload characteristics, including the main features of workload measurement data.
A concise description of considered dispatching policies is also given.
Numerical results of data-driven simulations for both artificial and experimental workloads are presented in \cref{sec:numres}.
Finally, conclusions are drawn in \cref{sec:conclusions}.

%

\section{System Model and Dispatching Policies}
\label{sec:modeldescription}

We consider $n$ identical \ac{FCFS} servers having processing speed $\mu$.
Jobs are offered to this server cluster with mean arrival rate $\lambda$ and size distributed according to a continuous random variable $S$ with \ac{CCDF} $G(x) = \mathcal{P}( S > x )$.
Service time, i.e., time required to complete a job on a server, is given by $X = S/\mu$.
The scale parameter $\mu$ is adjusted to impose that the coefficient of utilization $\rho$ averaged across servers has a desired value, namely, it is
\begin{equation}
\label{eq:rho}
\rho  = \frac{ \lambda \mathrm{E}[ S ] }{ n \mu }
\end{equation}

We also denote with $C_S$ the \ac{COV} of service times, defined as the ratio of the standard deviation of service time to the mean service time.
Jobs are instantaneously dispatched to one of the servers upon their arrival.
When dealing with experimental workload data, we account for the fact that a job can possibly be composed of several independent tasks. 
In that case, tasks get assigned individually to possibly different servers, according to the considered dispatching policy.
In any case, the reference metric is mean \emph{job} response time.
The job response time $R$ is defined as the time elapsing since job arrival until the last task of the job gets completed.
The \ac{MRT} is displayed either against $\rho$ or against the server cluster size $n$.
In either case, the overall computational power of the cluster, $n \mu$, is kept constant as $n$ varies, by adjusting $\mu$.
This makes the scaling of performance as $n$ varies non-trivial, given that the overall amount of computational resource is the same for each value of $n$, but the parallelism degree of the server cluster varies.

\subsection{Artificial workload}
To gather extensive performance evaluation data, we consider an M/G-type theoretical model setting.
We assume Poisson arrival of jobs, each one composed of a single task.
Job size is assumed to be independently generated for each job, drawn from the \ac{CCDF} $G(x)$.
In our numerical evaluation, we consider a Weibull distribution, i.e., 
\begin{equation}
G(x) = e^{ - \left( \frac{ x }{ a } \right)^b } \, , x > 0.
\end{equation}
The parameters $a$ and $b$ are found from the first two moments of job size, taking into account that 
\begin{align}
\label{}
    &\mathrm{E}[ S ] = a \Gamma\left( 1 + \frac{ 1 }{ b } \right) = \frac{ a }{ b } \Gamma\left( \frac{ 1 }{ b } \right)   \\
    &\mathrm{E}[ S^2 ] = \frac{  2 a^2 }{ b } \Gamma\left( \frac{ 2 }{ b } \right) 
\end{align}
where $\Gamma(z) = \int_{0}^{\infty}{ u^{z-1} e^{ - u } \, du }$ is the Euler Gamma function.

Specifically, we set $\mathrm{E}[ S ] = 1$ (so that the mean job service time on a standard server of speed $\mu = 1$ is the unit of measure of time), and $\mathrm{E}[ S^2 ] = \left( \mathrm{E}[ S ] \right)^2  \left( 1 + \text{\ac{COV}}^2 \right)$.

The mean arrival rate is fixed to achieve a desired load $\rho = \lambda \mathrm{E}[ S ]$, while the processing speed $\mu$ is set to $\mu = 1/n$ so that the overall computational budget of a server cluster made up of $n$ servers be fixed at 1.

\subsection{Experimental workload dataset}
\label{sec:dataset}

This section describes the Google traffic dataset released in 2020, capturing one month of user and developer activity in May 2019. 
It provides detailed resource usage data, such as \ac{CPU} and \ac{RAM}, across eight Google data centers ("Borg cells").

Traffic data includes five tables detailing users' resource requests, machine processing, and task evolution within Borg's scheduler. 
Users submit jobs comprising one or more tasks (instances), each requiring specific \ac{CPU} time and memory space. 
Resource units are based on \ac{GNCU} (default machine computational power).
Exposed \ac{CPU} time associated with the execution of a task assumes that servers are equipped with one \ac{GNCU}.

Jobs are composed of one or more tasks, each of which follows a life cycle.
Tasks are queued or processed based on load and classified upon completion as ``FINISH'' (success) or ``KILL,'' ``LOST,'' or ``FAIL'' (failure) \cite{wilkes2019clusterdata}. 
Only ``FINISH'' tasks were analyzed, as they provide reliable resource data. 
The dataset includes task identifiers, \ac{CPU} time (min, max, avg over 300-sec windows), and 1-sec samples per window of resource usage.

Digging into Google's data, the workloads have been reconstructed and the following key features are used in this work\footnote{The dataset for all eight data centers is available at \textcolor{blue}.{\url{https://github.com/MertYILDIZ19/Google_cluster_usage_traces_v3_all_cells}}}.
\begin{itemize}
	\item $J_j$ - number of tasks belonging to job $j$.
	\item $A_j$ - Arrival time (in seconds) of job $j$. All tasks belonging to the same job arrive at this time.
	\item $S_{ij}$ - \ac{CPU} time (in seconds) required to process task $i$ of job $j$ on a reference server equipped with one \ac{GNCU}.
\end{itemize}


The whole considered dataset (31 days worth of measured activity in Borg cell c) consists of 4399670 jobs, comprising 7010742 tasks.
Most jobs consist of a single task (96.6\%), yet ``monster'' jobs are also recorded, with the largest ones comprising up to 11160 tasks.
The mean job \ac{CPU} time is 10.83 s \footnote{We remind that all \ac{CPU} times are given with reference to a standard server equipped with 1 \ac{GNCU}.}.
Only 13.25\% of jobs require \ac{CPU} time larger than the mean, while 99\% of jobs have \ac{CPU} time less than 32.26 s, 99.9\% of jobs less than 452.95 s, and 99.99\% of jobs less than 4974.2 s.
The largest 0.1\% of jobs is responsible for 67.6\% of the overall required \ac{CPU} effort.
These few numbers give evidence of the extreme variability of workload described in Google's data.

The mean arrival rate of jobs $\lambda$ is estimated from data, as well as the job and task size distribution (used to set thresholds in some dispatching policies).
The server speed $\mu$ is adjusted for each given cardinality $n$ of the server cluster to achieve a desired server utilization coefficient $\rho$, according to \cref{eq:rho}.

\subsection{Dispatching policy}
\label{subsec:dispatchingpolicies}

For the ease of the reader, we provide a concise summary of the considered dispatching policies.

\subsubsection{Round Robin}
This policy does not rely on knowledge of either job size or server state, yet it exploits the history of decisions on job assignment.
The $k$-th arriving job is assigned to server $1 + (k - 1) \mod n$, for $k \ge 1$.

\subsubsection{Join Idle Queue}
This is an example of a policy that relies only on some server state information but does not require any information about job size.
The dispatching process maintains a look-up table of $n$ bits.
All bits are set to 1 initially.
Upon job arrival, the dispatcher picks a server at random among those having their bit at 1, sends the job to that server, and resets the corresponding bit in the table.
If no bit is at 1 in the table, the dispatcher sends the job to a server picked uniformly at random.
Whenever a server becomes idle (i.e., the last job with that server leaves the server, having been completed), the server sends a message to the dispatching process, so that the dispatching process can restore the server's bit at 1 in its table.

\subsubsection{Least Work Left}
This policy requires full knowledge of job size upon arrival, as well as keeping track of the current status of servers.
Let $W_j(t)$ denote the unfinished work at server $j$ at time $t$.
If servers are work-conserving, the dispatching process can keep track of $\mathbf{W}(t) = [W_1(t),\dots,W_n(t)]$ by knowing job sizes and the dispatching decision history.
Upon arrival of a job at time $a$, the job gets assigned to the server having minimum unfinished work, i.e., to server $j^* = \arg\min [W_1(a),\dots,W_n(a)]$.
Ties are broken at random.

\subsubsection{Multi-band flexible \ac{CARD}}
This is an example of dynamic policy exploiting both server status and job size knowledge.
Let thresholds $m_i$ and $c_i$ be defined as follows.
The thresholds $m_i$, $i = 1,\dots,n$, are defined so that the fraction of system load due to jobs with size falling in the interval $[m_i,m_{i+1})$ be equal to $1/n$ for $i = 1, 2, \dots, n-1$.
Moreover, we set $m_0 = 0$ and $m_{n+1} = \infty$.
Then, the fraction of system load due to jobs whose size is less than $m_1$ or larger than $m_n$ must be equal to $1/(2 n)$.
It is easy to verify that this is equivalent to setting $m_i$ so that the following equality holds
\begin{equation}
\frac{ 1 }{ \mathrm{E}[ S ] } \int_{0}^{m_i}{ x f(x) \, dx } = \frac{ i - 1/2 }{ n }
\end{equation}
for $i = 1,\dots,n$, where $f(x)$ is the \ac{PDF} of job sizes.
As for the thresholds $c_i$, Authors of \cite{Xie2024} suggest to set $c_i = m_i / \sqrt{ 1 - \rho }$, for $i = 1,\dots,n-1$.

Then, the \ac{CARD} policy works as follows.
Upon job arrival at time $a$, unfinished work levels $W_i(a), \; i = 1,\dots,n$, at servers are sorted in ascending order.
Let $W_{(i)}(a)$ denote the ordered sequence of unfinished work levels, i.e., $W_{(1)}(a) \le W_{(2)}(a) \le \dots \le W_{(n)}(a)$.
Let also $K_{(i)}(a)$ denote the index of the $i$-th ranked server, i.e., the server having the $i$-th smallest unfinished work amount.
Let $s$ denote the size of the newly arrived job.
\begin{enumerate}
  \item If $s < m_1$, the job gets assigned to server $K_{(1)}(a)$.
  \item if $s > m_n$, the job gets assigned to server $K_{(n)}(a)$.
  \item If $s \in [m_i,m_{i+1})$, the job gets assigned to server $K_{(i)}(a)$ if $W_{(i)}(a) \le c_i$, to server $K_{(i+1)}(a)$ otherwise ($i = 1,\dots,n-1$).
\end{enumerate}

\subsubsection{Two-stage dispatching}
The two-stage dispatching architecture was first introduced in~\cite{yildiz2025meritsimplepoliciesbuying}, further refined in~\cite{yildiz2025dispatchingodysseyexploringperformance}, and in this paper, we generalize and apply its most effective variant across different dispatching policies. The mechanism operates as follows: the servers are divided into two sub-clusters (or stages) of sizes \(n_1\) and \(n_2 = n - n_1\). Upon arrival, a job is dispatched to one of the \(n_1\) servers in the first stage according to a chosen dispatching policy \(D\) (e.g., \ac{RR}, \ac{JIQ}, or \ac{LWL}). 
If the job is not complete within a time interval of \(\theta/\mu\) (i.e., if its size exceeds \(\theta\)), it is preempted and transferred to the second stage, where it is assigned to one of the \(n_2\) servers using the same policy \(D\) and then restarted to be completed in the second stage. 
All servers operate under \ac{FCFS} scheduling. 
The only additional mechanism is a timer on the first-stage servers; if the timer expires, the job currently running is halted and moved to the second stage.

Notably, when using \ac{RR}, this policy requires neither knowledge of the job size nor monitoring of server status during dispatching. 
In the case of \ac{JIQ}, only the idle/busy state of each server is needed, a requirement that is straightforward to implement in realistic systems.
Its performance obviously depends on the two meta-parameters $n_1$ and $\theta$.
In this work we consider a static optimization of those parameters, leaving for further work the definition of an algorithm for automatic adaptation of those parameters.

%

\section{Numerical Results}
\label{sec:numres}

In this section, we evaluate the performance of various dispatching policies through simulations. 
Our objective is to address the following key questions:

\begin{itemize}
    \item How does the degree of parallelism (i.e., the number of servers) affect performance, and what is the optimal server range for performance optimization?
    \item How does the Two-Stage dispatching policy compare to the traditional Single-Stage dispatching and the size and state-aware, heavy-traffic optimal policy, \ac{CARD}?
    \item How do the Two-Stage and \ac{CARD} policies perform under Poisson arrivals and Weibull-distributed service times compared to real-world Google workload traces?
\end{itemize}

As discussed in the preceding sections, we employ four benchmark dispatching policies: \ac{RR}, \ac{JIQ}, \ac{LWL}, and \ac{CARD}. 
Although dispatching policies have been extensively studied, we selected these due to their varying levels of sophistication. 
Specifically, \ac{RR} serves as a simple, stateless baseline; \ac{JIQ} is a stateful, but size-agnostic policy; and \ac{LWL} is the most sophisticated of the three, incorporating both size and state information. 
We also include \ac{CARD} in our evaluation because it is provably optimal~\cite{Xie2024} in heavy-traffic regimes for size- and state-aware dispatching. 
The point we aim to make is that a significant performance gain can be reaped by changing the server cluster architecture from single-stage to two-stage.
Therefore, we apply \ac{RR}, \ac{JIQ}, and \ac{LWL} dispatching in a two-stage framework and evaluate their performance accordingly.

Heavy-tailed distributions are prevalent in computer systems and networks (e.g., \cite{mahanti2013}), and their tendency to result in high mean response times underscores the importance of an effective dispatching policy. 
To reproduce this variability we consider artificially generated workloads in \cref{subsec:weibullworkload} and real-world workloads obtained from Google measurements in \cref{subsec:googleworkload}.

\subsection{Performance under Weibull-Distributed Workload}
\label{subsec:weibullworkload}
In this Section, we first evaluate the system under an M/G-type model in which job sizes follow a Weibull distribution with a mean normalized to 1. 
To capture both moderate- and heavy-variability regimes, we set the \ac{COV} to either 1 or 10. 
We compare the mean response time under different dispatching policies by normalizing it with respect to the theoretical mean response time of an equivalent single-server \ac{FCFS} M/G/1 queue, $E[R_{M/G/1}]$. 
Our goal is to examine how these policies perform as we vary both the number $n$ of servers and the system load $\rho$, while keeping total processing capacity fixed.

\begin{figure}[]
\centering
\includegraphics[width=0.7\columnwidth]{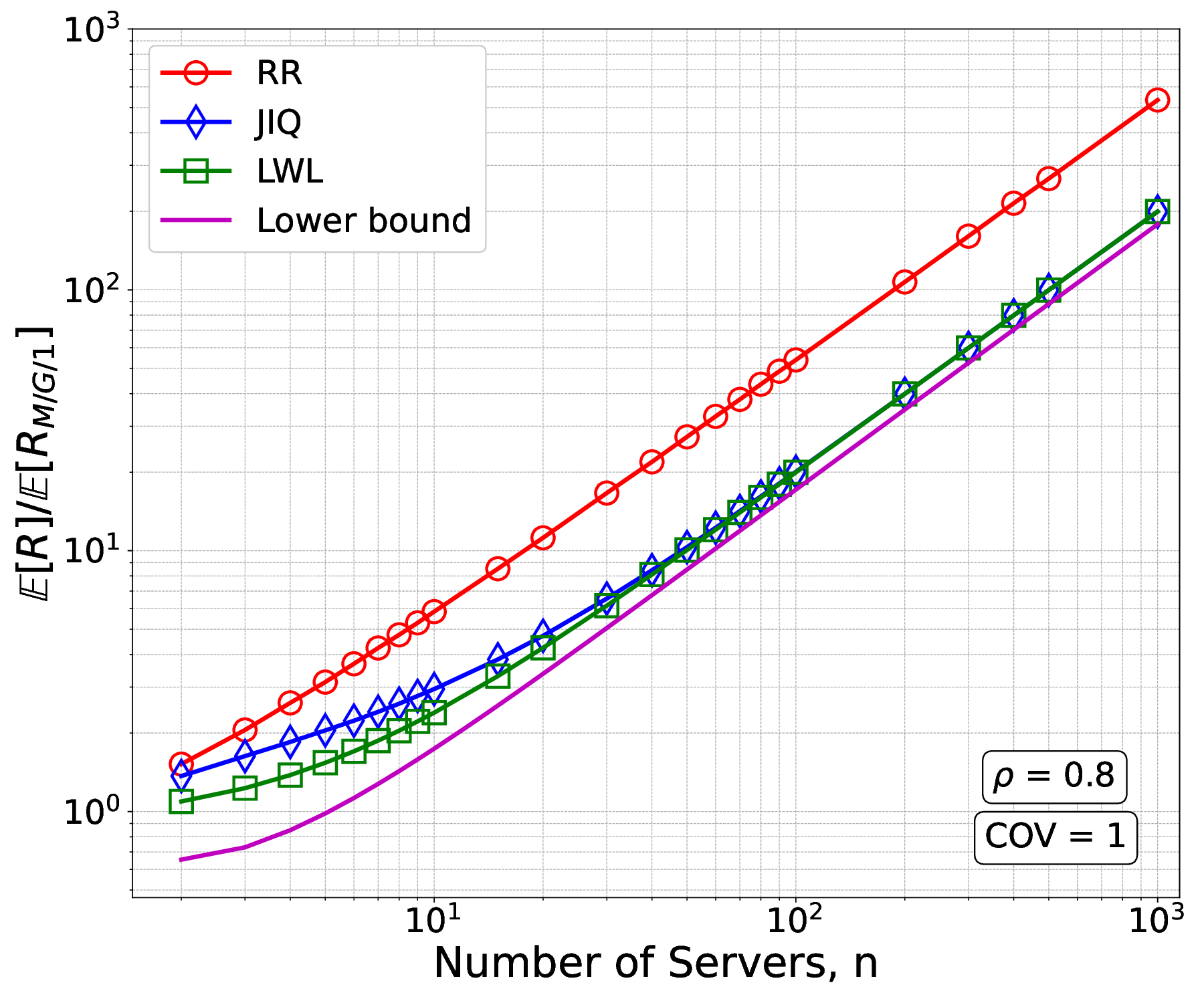}
\caption{\ac{MRT} as a function of the number of servers. \textbf{Note:} \ac{COV} = 1.}
\label{fig:ER_Nse_Mod_Sim_COV1}

\vspace{-0.325cm}
\end{figure}

\begin{figure}[]
\centering
\includegraphics[width=0.7\columnwidth]{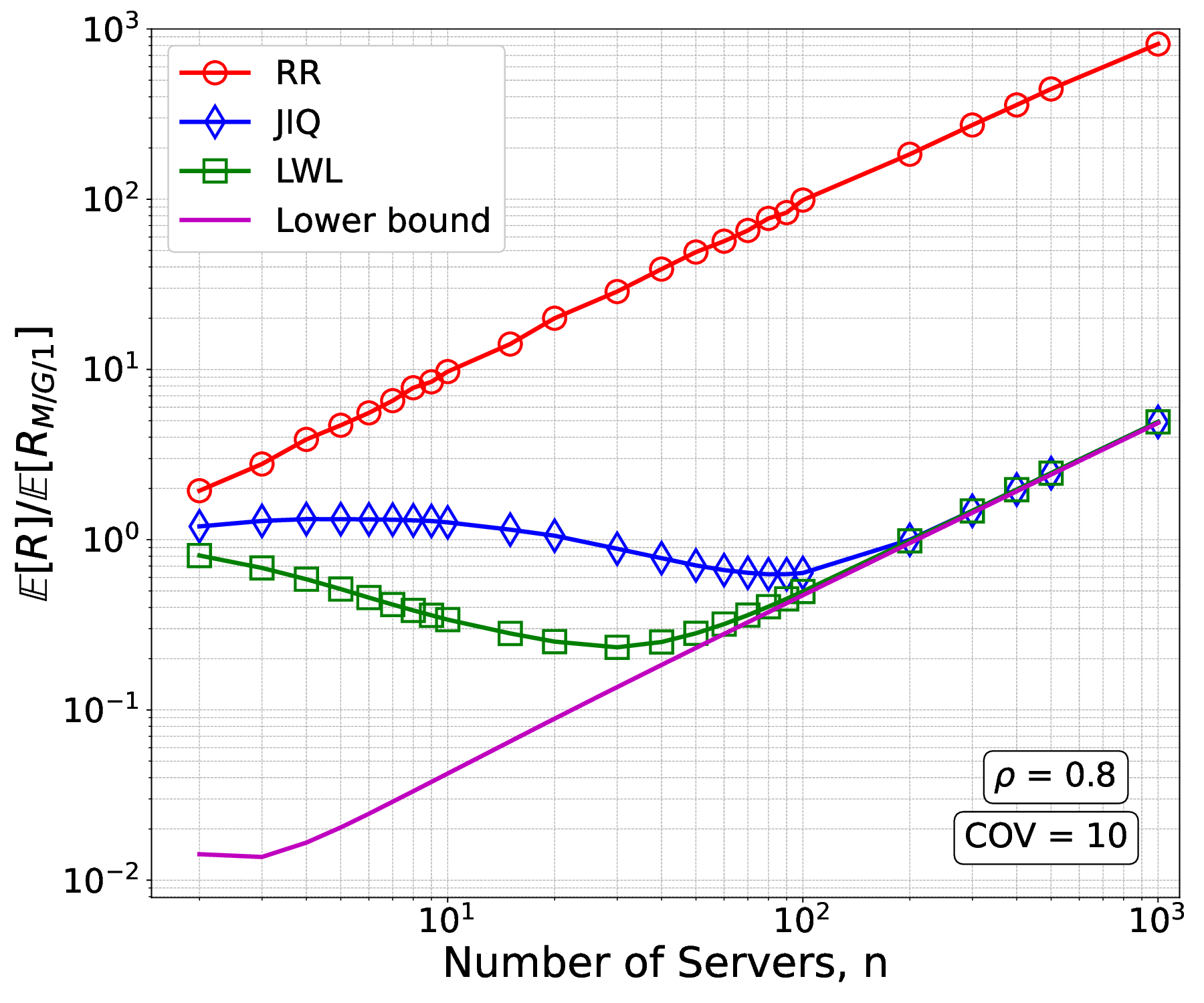}
\caption{\ac{MRT} as a function of the number of servers. \textbf{Note:} \ac{COV} = 10.}
\label{fig:ER_Nse_Mod_Sim_COV10}

\vspace{-0.325cm}
\end{figure}

First, the effect of the server cluster size is evaluated.
We will show that as $n$ grows, near-optimal performance can be achieved even with relatively simple policies, while for low or moderate values of $n$ smart dispatching policies have their merit.
\cref{fig:ER_Nse_Mod_Sim_COV1} and \cref{fig:ER_Nse_Mod_Sim_COV10} plot the normalized mean response time as a function of the number of servers for \ac{COV} equal to 1 and 10, respectively, for the three basic dispatching policies \ac{RR}, \ac{JIQ}, and \ac{LWL}, with $\rho = 0.8$.
The lower bound derived in \cite{Xie2024} is shown as well. 

These results offer some key insights. 
When \(n\) becomes large enough, both \ac{JIQ} and \ac{LWL} perform almost as well as the theoretical lower bound.
On the contrary, \ac{RR} exhibits an offset with respect to the lower bound, the bigger the higher the \ac{COV}.
Hence, \emph{for large parallelism degree of the cluster server, there is no need to resort to size-aware dispatching policies}.
Basic state information (namely, idle/busy server state) is enough to reap near-optimal mean response time.
This is not a trivial result, given that the overall computational power of the server cluster is fixed to constant values as $n$ varies, by scaling the server speed $\mu$.
Therefore, service time grows as $n$ grows.


\textit{A significant offset with respect to optimal performance arises at moderate levels of $n$}. 
For \(n \in \{2, \dots, 100\}\), \ac{JIQ} and \ac{LWL} consistently outperform \ac{RR} for both \(\mathrm{\ac{COV}}=1\) (negative exponential distribution) and \(\mathrm{\ac{COV}}=10\).
In the exponential regime (\(\mathrm{\ac{COV}}=1\)), \ac{JIQ} and \ac{LWL} yield almost the same mean response times, indicating that knowing job sizes (\ac{LWL}) offers a minimal advantage over purely state-based dispatching (\ac{JIQ}). 
By contrast, under the heavy-tailed regime (\(\mathrm{\ac{COV}}=10\)), the impact of large jobs is magnified, making any policy that leverages state information substantially more effective than the baseline \ac{RR}.
From these observations, we conclude that the most pronounced performance distinctions emerge in moderate-scale systems, between a few to a few hundred servers. 
Beyond a sufficiently high degree of parallelism, both \ac{JIQ} and \ac{LWL}  closely match the lower bound, rendering the choice of dispatching policy less critical.
Consequently, evaluation and optimization will focus on the range \(2 \leq n \leq 100\), where differences between dispatching strategies are most apparent.
Within this critical region, we compare the performance of both single-stage traditional and \ac{CARD} with two-stage dispatching policies.

Before proceeding, we clarify the two-stage setting as follows.

\textbf{Two-stage Architecture and Threshold Selection:} 
For the two-stage variants of \ac{RR}, \ac{JIQ}, and \ac{LWL}, we optimize two key parameters: (i) the threshold \(\theta\), selected from a set of quantiles of the job size distribution, and (ii) the first-stage size \(n_1\), with the second stage comprising $n_2 = n - n_1$ servers. 
By testing various quantile-based thresholds and different splits of \(n_1\) and \(n_2\), we identify an effective boundary for offloading jobs from the first to the second stage. 
In practice, this approach significantly enhances performance for shorter jobs, as it isolates heavy jobs onto the second stage while allowing shorter jobs to be completed rapidly in the first stage.

\begin{figure}[]
\centering
\includegraphics[width=0.7\columnwidth]{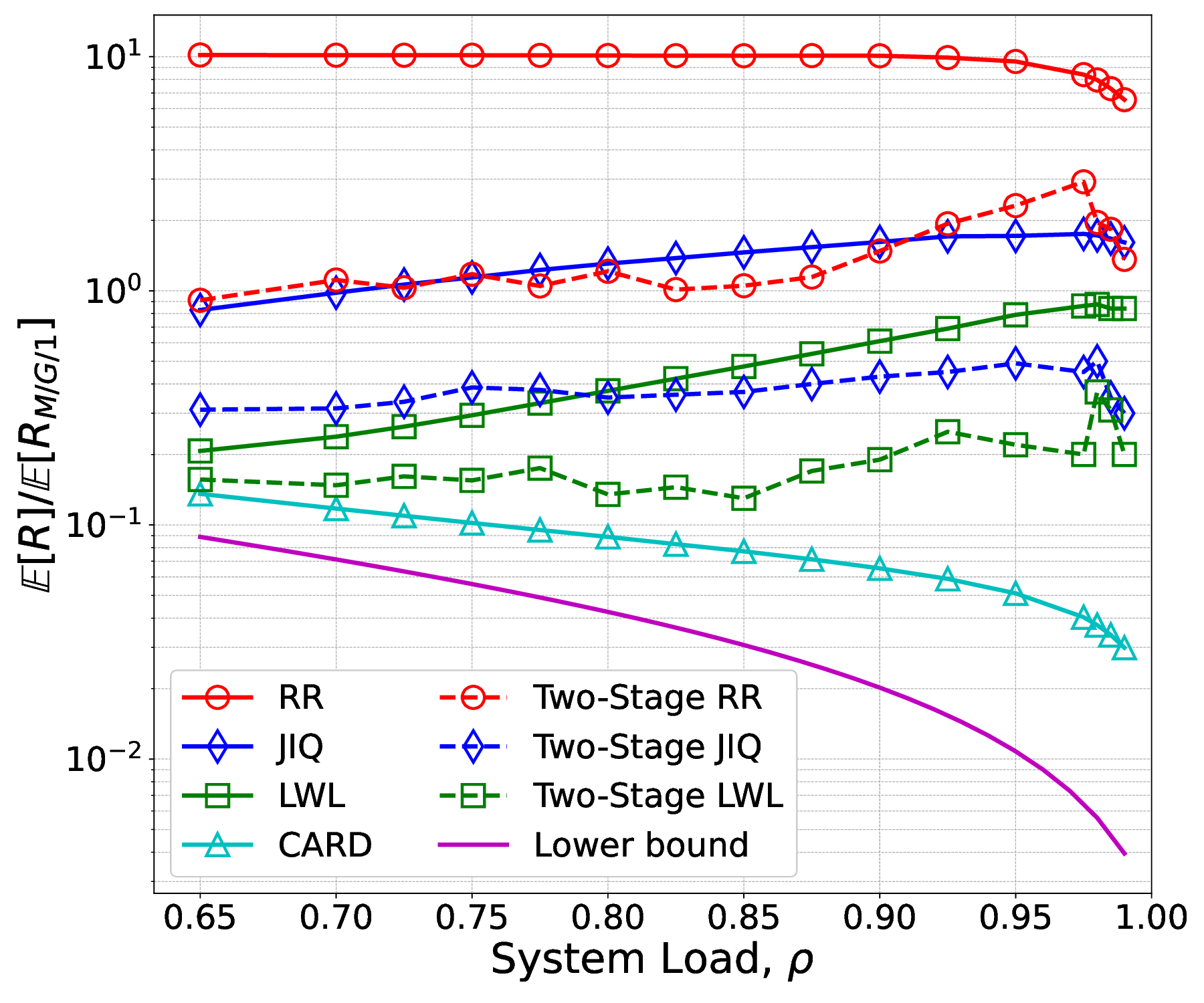}
\caption{\ac{MRT} as a function of the system load. \textbf{Note:} \ac{COV} = 10 and n = 10.}
\label{fig:ER_Nse_Weibull}

\vspace{-0.325cm}
\end{figure}

\Cref{fig:ER_Nse_Weibull} shows the normalized mean response time, \(\mathbb{E}[R] / \mathbb{E}[R_{M/G/1}]\), as a function of the system load \(\rho\) setting the number of server n = 10. 
One can see that each policy’s performance varies smoothly with \(\rho\), with \ac{RR} (red) generally yielding the highest mean response times at moderate to high loads, and \ac{CARD} (cyan) consistently remaining the best-performing single-stage policy across the entire load range. \Cref{fig:ER_Nse_Weibull} illustrates that the optimized two-stage \ac{RR} variant (dashed red) significantly narrows the gap between single-stage \ac{RR} and more sophisticated policies, highlighting how selectively separating large jobs can substantially improve overall performance. 
Nevertheless, at extremely high loads close to \(\rho = 1\), the advantage of knowing job sizes (\ac{LWL}) or at least the state of the server (\ac{JIQ}) still emerges in slightly lower response times. 

\begin{figure}[]
\centering
\includegraphics[width=0.7\columnwidth]{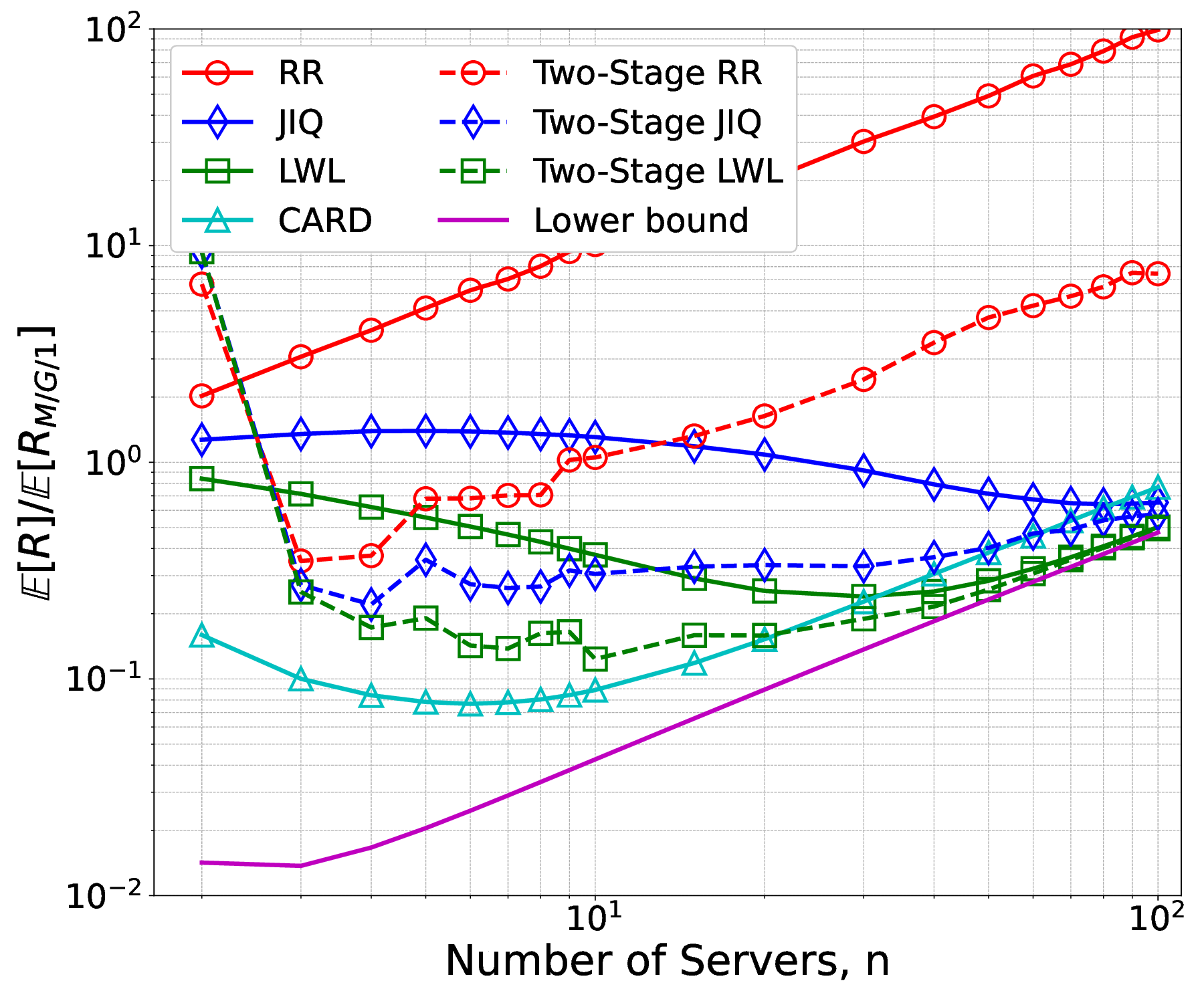}
\caption{\ac{MRT} as a function of the number of servers. \textbf{Note:} \ac{COV} = 10 and $\rho = 0.8$.}
\label{fig:ER_Rho_Weibull}

\vspace{-0.325cm}
\end{figure}

In \Cref{fig:ER_Rho_Weibull}, we fix \(\rho = 0.8\) and vary the number of servers \(n\) to clearly illustrate the impact of server parallelism.
When \(n\) is small, the two-stage policies perform poorly; however, increasing the server count to around 10--20 dramatically reduces the mean response time. 
In realistic scenarios, where one does not employ an extremely small number of servers (which would require each server to be exceptionally powerful to maintain a constant total capacity), the \ac{RR} policy, in particular, fails to achieve satisfactory performance.
\ac{LWL} and \ac{JIQ}, by contrast, already achieve good performance for small \(n\) and then see diminishing returns beyond moderate-scale configurations, especially as the cluster size grows further. 
Eventually, when \(n\) is very large, the lines tend to converge, indicating that high parallelism weakens the distinction among policies. 
In all configurations, \ac{CARD} continues to demonstrate the best single-stage performance, while the threshold-optimized two-stage algorithms remain close behind, underscoring the effectiveness of migrating heavier jobs to a second stage to shield latency-sensitive, shorter jobs from undue delays.

\subsection{Performance under Real Google Workload}
\label{subsec:googleworkload}
We now turn to a real-world Google data center workload that encompasses a full day of measured user activity. 
Unlike the artificially generated workload, jobs in this dataset might comprise multiple tasks, each with its own CPU requirement (i.e., task size). 
An in-depth analysis of Google workload data shows that \cite{yildiz2024}: (i) all tasks belonging to the same job arrive at the same time (within the accuracy of timestamps); (ii) tasks belonging to the same job appear to be independent, i.e., there is no evidence of the need for completing a given task to start running another task of the same job.
Hence, in our simulations, tasks are dispatched individually according to the selected policy, i.e., \ac{RR}, \ac{JIQ}, \ac{LWL}, \ac{CARD}, and their two-stage variants. 
Mean response time is measured as \emph{job} completion time, i.e., the elapsed time since the job's arrival,  until \emph{all} of its tasks have finished.

\begin{figure}[]
\centering
\includegraphics[width=0.7\columnwidth]{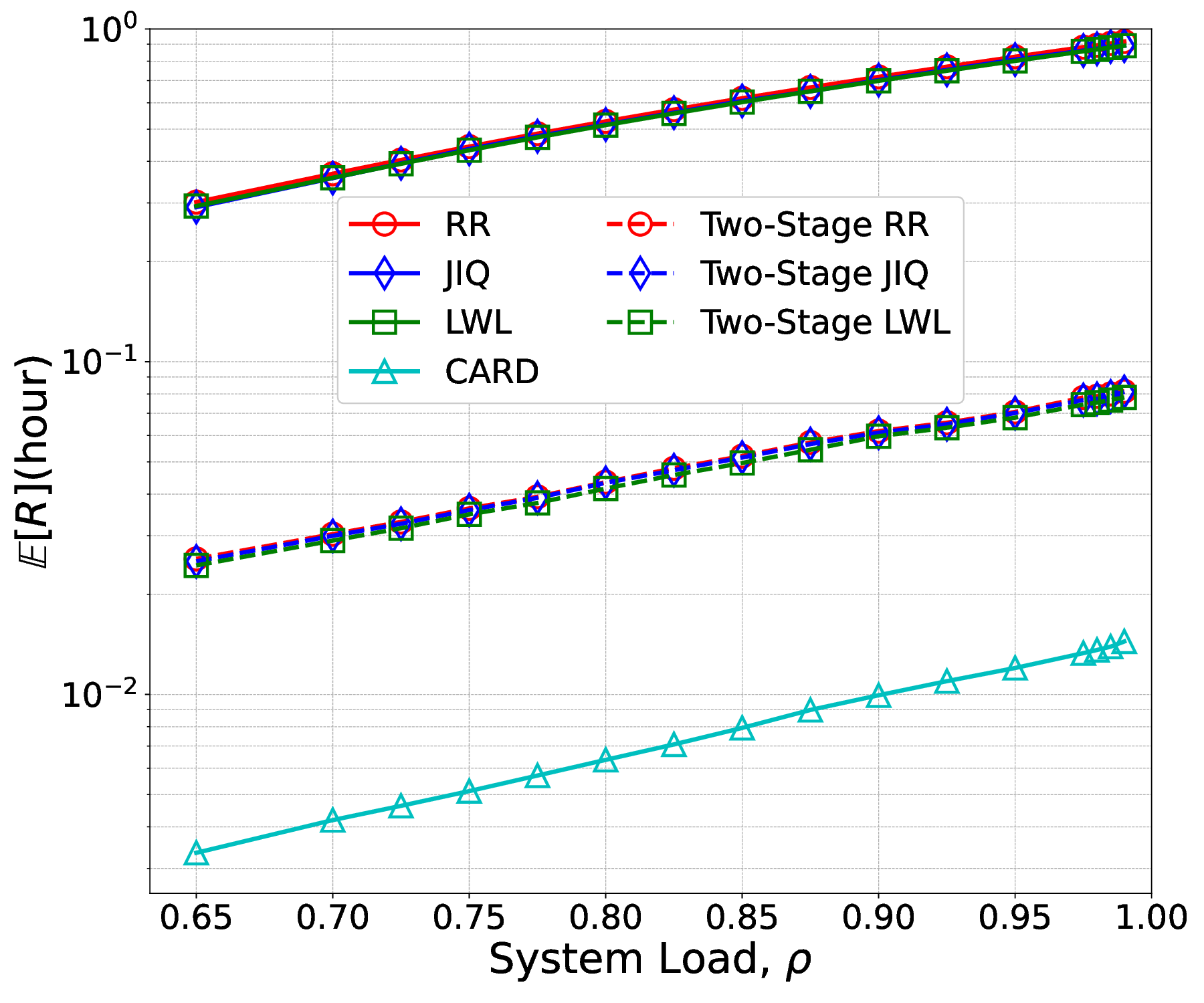}
\caption{\ac{MRT} as a function of the system load. Google data has been used including the task-level data structure. \textbf{Note:} $n = 10$}
\label{fig:Google_ER_Rho}
\vspace{-0.325cm}
\end{figure}

\begin{figure}[]
\centering
\includegraphics[width=0.7\columnwidth]{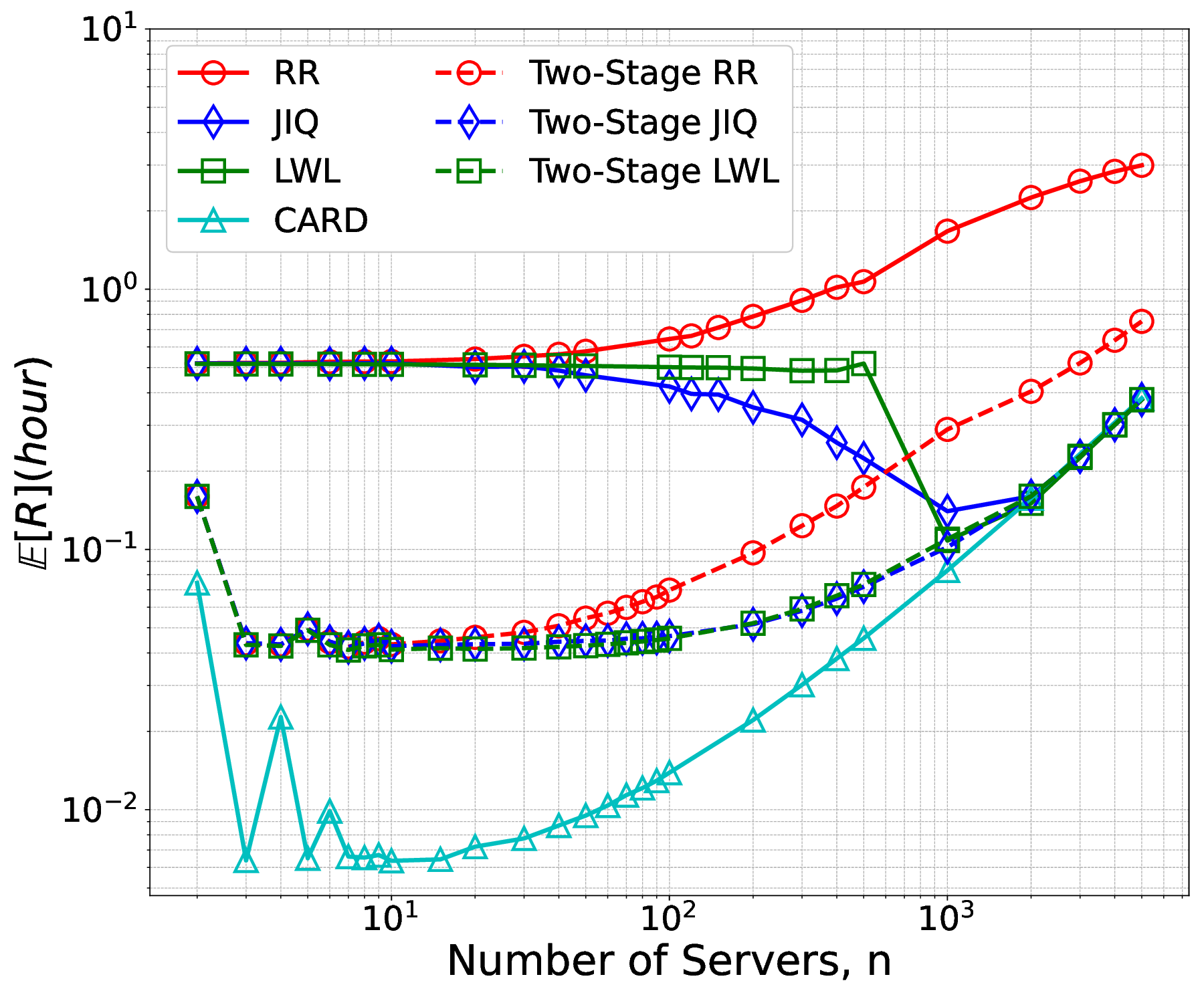}
\caption{\ac{MRT} as a function of the number of servers. Google data has been used including the task-level data structure. \textbf{Note:} $\rho = 0.8$}
\label{fig:Google_ER_Nse}

\vspace{-0.325cm}
\end{figure}

\Cref{fig:Google_ER_Rho} and \Cref{fig:Google_ER_Nse} plot the mean job response time (in hours) as we vary the system load $\rho$ for $n = 10$, and the cluster size $n$ for $\rho = 0.8$, respectively. 

Focusing first on \Cref{fig:Google_ER_Rho}, it is apparent that exploiting state or size information makes little difference among \ac{RR}, \ac{JIQ}, and \ac{LWL}, while changing the architecture of the server cluster from one to two-stage brings about a major improvement of mean response time performance.
\ac{CARD} (cyan curves) consistently outperforms the other policies, thus showing that state and size information \emph{can} make a difference if played in the right way.
It is however remarkable that a good part of the performance advantage gained by \ac{CARD} over single-stage policies can be reaped by adopting the baseline \ac{RR} policy but moving to two-stage architecture.

Turning to \Cref{fig:Google_ER_Nse} we observe that \ac{CARD} still boasts a noticeable performance edge, but two-stage \ac{JIQ} and \ac{LWL} (dashed blue and green curves) remain close to \ac{CARD} and clearly outperform their single-stage counterparts. 
This shows that carefully splitting the cluster can substantially improve \ac{JIQ} and \ac{LWL}, especially under moderate to large values of $n$. 
It is also notable that the two-stage \ac{LWL} policy offers no performance advantage over the simpler two-stage \ac{JIQ}. In the single-stage setting, \ac{JIQ} even outperforms \ac{LWL} in some cases, highlighting that policies leveraging more information—such as job size—do not necessarily yield better performance than simpler, state-based alternatives.

In this case, \ac{LWL} fails to provide some imbalance among different servers, as achieved naturally by \ac{JIQ} due to randomization of task assignments when all servers are busy.

Overall, these results confirm that real data center workloads are heavily skewed in both job sizes and concurrency levels. 
While \ac{CARD} leverages both queue state and job-size knowledge to achieve robust performance, the two-stage approach continues to offer significant improvements for \ac{RR} and yields competitive outcomes for \ac{JIQ} and \ac{LWL}. 
This indicates that architectural modifications can be a powerful alternative to deploying more complex, information-intensive dispatching algorithms when tackling real, large-scale workloads.

%

\section{Conclusions}
\label{sec:conclusions}

In this paper, we demonstrated that a two-stage architecture can closely approach the performance of advanced size- and state-aware policies in \ac{FCFS} server clusters. By introducing a service-time threshold and partitioning servers into two stages, even simple policies like \ac{RR} exhibit substantial gains—especially under heavy-tailed workloads, where isolating large jobs shields short ones from excessive delays.

Experiments on synthetic Weibull workloads and real Google traces confirm that \ac{CARD}, with full size and state-awareness, remains the top performer. However, two-stage variants of \ac{JIQ} and \ac{LWL} often achieve comparable \ac{MRT} with reduced implementation complexity and limited coordination, offering a practical alternative for large-scale systems with high job size variability.

For future work, we aim to design an adaptive mechanism that continuously learns the optimal threshold by tracking service times of arriving jobs. This would enable the system to respond dynamically to workload shifts and better isolate large jobs. We also plan to investigate a two-stage version of \ac{CARD}, to determine whether it can surpass its single-stage performance. Finally, we intend to quantify the overhead of transferring jobs between stages, which was not addressed in this study but may impact performance under high load or resource contention.

%
%

\section*{Acknowledgment}
This work was supported by the European Union - Next Generation EU under the Italian National Recovery and Resilience Plan (NRRP), Mission 4, Component 2, Investment 1.3, CUP B53C22004050001, partnership on “Telecommunications of the Future” (PE00000001 - program “RESTART”)

\bibliographystyle{IEEEtran}
\bibliography{references}

\end{document}